# Bridging the Early Science Gap with Artificial Intelligence

Evaluating Large Language Models as Tools for Early Childhood Science Education

Annika Bush
Research Center Trustworthy Data Science and Security,
University Alliance Ruhr
TU Dortmund
Dortmund, Germany
annika.bush@tu-dortmund.de

Amin Alibakhshi
Research Center Trustworthy Data Science and Security,
University Alliance Ruhr
Ruhr University Bochum
Bochum, Germany
amin.alibakhshi@rub.de

## Abstract

Early childhood science education is crucial for developing scientific literacy, yet translating complex scientific concepts into age-appropriate content remains challenging for educators. Our study evaluates four leading Large Language Models (LLMs) - GPT-4, Claude, Gemini, and Llama - on their ability to generate preschool-appropriate scientific explanations across biology, chemistry, and physics. Through systematic evaluation by 30 nursery teachers using established pedagogical criteria, we identify significant differences in the models' capabilities to create engaging, accurate, and developmentally appropriate content. Unexpectedly, Claude outperformed other models, particularly in biological topics, while all LLMs struggled with abstract chemical concepts. Our findings provide practical insights for educators leveraging AI in early science education and offer guidance for developers working to enhance LLMs' educational applications. The results highlight the potential and current limitations of using LLMs to bridge the early science literacy gap.

## 1 Introduction

Science education in early childhood has gained increasing recognition as a crucial foundation for children's cognitive development and future learning. However, already in preschool age, there are significant gaps between children when it comes to science literacy as the competence to understand core disciplinary ideas and practices [3]. Research shows that primary school cannot bridge this early developed gap which makes it even more important to teach kids basic science in kindergarten already [14]. While young children exhibit natural curiosity about scientific phenomena, many educators actively seek additional resources and support to optimize their presentation of scientific concepts to young learners [12].

Rapid advances in Artificial Intelligence (AI), particularly Large Language Models (LLMs), offer promising opportunities to address this challenge by automatically transforming complex scientific content into child-friendly explanations. Recent developments in LLM technology have demonstrated remarkable capabilities in content generation and adaptation [15]. These models can now process and reformulate complex information, making them potentially valuable tools for creating educational content [9]. However, their effectiveness in generating content specifically tailored to very young children, particularly preschoolers, remains largely unexplored. This gap is particularly significant given the unique linguistic and cognitive needs of preschool children, who are just beginning to develop their understanding of the world around them.

While several LLMs are currently available, each with its own characteristics and capabilities, there has been no systematic evaluation of their suitability for generating scientific content for early childhood education. This study addresses this research gap by examining which LLM is most suitable for generating child-appropriate scientific content, focusing specifically on four-year-old children's comprehension needs. We chose four popular LLMs from OpenAI (GPT), Meta (Llama), Google (Gemini), and Anthropic (Claude). Our research focuses on the three fundamental natural science disciplines: biology, chemistry, and physics.

By evaluating LLM-generated content across these disciplines, we aim to provide insights that can benefit both educators and content developers in the field of early childhood science education (ECSE). Moreover, it contributes to our understanding of LLMs' capabilities in educational content adaptation, particularly for young learners. The involvement of experienced nursery teachers as evaluators ensures that our findings are grounded in practical educational expertise.

## 2 Related Work

### 2.1 LLMs in Educational Content Generation

Recent developments in LLM technology have shown promising results in content adaptation across various domains. Several studies have explored LLMs' capabilities in simplifying complex texts [2, 21] and generating educational content [1, 13].

However, the specific application of LLMs in creating content for very young children remains largely unexplored, with only a few studies addressing this particular use case. A recent study by Bhandari and Brennan [5] used Meta's LLMs to assess the trustworthiness of AI-generated children's stories and evaluated the generated texts. They revealed that LLMs face challenges in generating children's stories that match the quality and nuance of human-written tales. However, a contrasting finding emerged from the research conducted by Weber et al. [20], which showed that parents perceived AI-generated stories as engaging, age-appropriate, and educational. It is important to note that Weber et al. [20] pursued a different research objective and worked with a different LLM. They used OpenAI's GPT-3 to analyze whether generated texts can facilitate vocabulary learning. Nonetheless, the disparity in both study findings highlights the complex nature of child-appropriate text generation, which is influenced by both the technical limitations of LLMs and differences in the perceived value of the content.

However, most research focusing on LLMs in education has focused on students in primary and secondary school or general audiences [9]. Additionally, over 90% of the studies focusing on LLMs in an educational context solely use and assess OpenAI's GPT models and their capabilities [9]. To our knowledge, no studies have focused on using LLMs to generate ECSE content.

### 2.2 Science Education in Early Childhood

Recent research has emphasized the importance of introducing scientific concepts during early childhood. Studies have shown that children as young as four can grasp basic scientific concepts when presented appropriately [17, 18]. While traditional approaches focus on hands-on experimentation and observation, the role of age-appropriate explanations has been increasingly recognized as crucial for building scientific understanding [19]. However, creating such explanations presents significant challenges, as it requires both scientific accuracy and alignment with young children's cognitive development stages.

### 2.3 Characteristics of Child-Appropriate Scientific Content

The literature identifies several key characteristics to assess whether texts are suitable for young children. These characteristics apply to texts in general and scientific content in particular. One of the most important aspects is simple and comprehensible age-appropriate language with short sentences [8, 22]. Research emphasizes the importance of sparking children's interest, curiosity, and interest [8, 22]. Additionally, studies have shown that four-year-olds particularly benefit from concrete explanations that relate to their immediate environment and experiences [12]. Additionally, a suitable length and complexity of explanations for this age group have been established through various educational studies [8]. Studies involving expert educators in content evaluation have proven particularly valuable, as they combine theoretical frameworks with practical teaching experience [8].

### 2.4 Research Gap and Hypotheses

While existing literature provides valuable insights into both ECSE and LLM capabilities, there is a notable gap in research specifically examining LLMs' effectiveness in generating scientific content for young children. Previous studies have either focused on content generation for older age groups or examined ECSE without considering AI-generated content. Our study aims to bridge this gap by systematically evaluating different LLMs' capabilities in generating age-appropriate scientific content for four-year-olds.

Since research on LLMs in educational contexts has predominantly focused on OpenAI's GPT models [9] and positive outcomes have been demonstrated [20], our first hypothesis is as follows:

H1: GPT-4 will perform better than other LLMs.

The ability of young children to grasp scientific concepts varies significantly based on the subject matter's abstractness and their ability to connect it to their immediate environment and experiences [12]. Different scientific disciplines inherently present varying levels of complexity and abstraction, which may pose distinct challenges for LLMs in generating age-appropriate explanations. Therefore:

H2: There will be significant differences in nursery teachers' ratings of LLM-generated content across scientific disciplines (biology, chemistry, and physics), with some disciplines being evaluated more favorably than others.

Research has established several distinct characteristics that determine the suitability of scientific content for young children, including comprehensible age-appropriate language, connection to children's immediate environment, and the ability to spark curiosity and interest [8, 22]. As these criteria represent fundamentally different aspects of child-appropriate content, and LLMs may have varying strengths in different aspects of content generation:

H3: There will be significant differences in nursery teachers' preferences for LLMs across the four evaluation criteria (comprehensibility, language, interest generation, and real-life relatedness).

## 3 Study Design

To investigate which LLM is most suitable for generating child-appropriate scientific content, we focused on the natural sciences of biology, chemistry, and physics. We generated texts using different LLMs and had them rated by nursery teachers as experts in the field.

In the first step, we decided on four topics per discipline based on compendiums in biology [7], chemistry [6], and physics [11] (see tab. 1). Based on these topics, we extracted one text per topic (12 texts in total) from the compendium.

Afterward, we chose four LLMs based on popularity and user interface (see tab. 2). The selection criteria focused on easily accessible platforms, as our study aimed to simulate realistic usage scenarios for nursery teachers. Since most educators lack specialized training in prompt engineering, we wanted our findings to be applicable in real-world settings where teachers might use simple prompts with these LLMs. This approach ensures that our results

**Table 1: Chosen topics by discipline**

| Biology | Physics | Chemistry |
|---|---|---|
| Digestion | Motion | Chemical reactions |
| Viruses | Energy | Ionic bonding |
| Photosynthesis | Electricity | Periodic table |
| Human brain | Atoms | Carbon |

remain relevant for practitioners without technical expertise in AI interaction.

**Table 2: Used Large Language Models**

| Company | LLM | Version |
|---|---|---|
| OpenAI | GPT | 4o |
| Anthropic | Claude | 3.5 Sonnet |
| Google | Gemini | 1.5 Flash |
| Meta | Llama | 3.1 |

The next phase of our research focused on prompt engineering. We conducted extensive testing with various prompts across the selected LLMs to determine the most effective approach. Through iterative refinement and comparison, we established a single standardized prompt: "transform the text for a 4-year-old child in 100 words max." This short prompt has been chosen to ensure consistency, minimize hallucinations (incorrect or misleading AI outputs), and to keep the length appropriate for young children's attention spans. Also we decided against longer prompts or few-shot prompting to mimic the realistic usage of LLMs by nursery teachers.

Using this optimized prompt, we proceeded to generate content with all four LLMs. For each of the twelve scientific topics, we generated texts in both English and German, resulting in 96 texts in total (12 topics×4 LLMs, each in two languages). The German versions were necessary for evaluation by German nursery teachers, while the English versions provide a valuable dataset for potential comparative studies in other countries and languages.

To establish a structured evaluation framework, we identified four key characteristics based on our literature review and pretesting, which included non-structured talks with nursery teachers: (1) child-appropriate language, (2) relation to children's everyday life, (3) comprehensibility for 4-year-olds, and (4) capacity to arouse interest. These characteristics served as our primary assessment criteria for analyzing the quality and appropriateness of the generated texts. All four criteria were rated on a 5-point Likert scale for each text. Additionally, the nursery teachers were asked to indicate their favorite text for each topic (one out of four).

The final evaluation phase for this late-breaking work involved 30 nursery teachers who served as expert evaluators. These professionals were asked to assess a total of 48 German texts, comprising the different versions generated for each original scientific text. The nursery teachers were recruited from 6 different urban early childhood education centers. All participants were professionally trained and certified according to national standards for early childhood education. Regarding technology familiarity, some teachers had limited experience with generative AI tools like ChatGPT for personal purposes (such as requesting recipes), but none had previously used AI technologies for professional educational purposes.

Their expertise in early childhood education provided valuable insights into the appropriateness and effectiveness of the generated content. Based on their comprehensive ratings, we were able to determine which LLM demonstrated the greatest proficiency in creating scientific content that was both accurate and accessible to four-year-old children. Additionally, we were able to identify the strengths and weaknesses of the different LLMs when it comes to specific aspects of generating child-appropriate content, like language or comprehensibility.

## 4 Results
### 4.1 Examples of LLM-Generated Content

To illustrate the differences between outputs from different LLMs, we present one example of generated content for the same scientific concept across the four models.

Example topic: *Biology - Viruses* [7].

OpenAI GPT-4o: "A virus is a tiny thing, much smaller than a cell, made of just a few parts, like a coat (cover) and a small bit of genetic instructions (kind of like a recipe). Viruses need to go inside a living cell to grow because they can't do it by themselves. They "borrow" parts from the cell to make more viruses. Sometimes, this makes the cell sick or even causes it to break open. Scientists learned about viruses by studying plants that got sick, discovering that tiny, invisible things (viruses) were the cause."

Anthropic Claude Sonnet 3.5: "Sometimes tiny things called viruses can make us feel sick, just like when you get a cold. Imagine viruses are like tiny puzzle pieces that are so small you can't see them! They're smaller than even the tiniest ant. These little pieces can't do anything on their own - they need to find a home inside other living things, like people, animals, or plants, to grow and make more of themselves. It's a bit like how you need a house to live in. That's why we wash our hands and take medicine when we're sick - to keep these tiny visitors away!"

Google Gemini 1.5 Flash: "Imagine tiny ninjas! These ninjas are so small you can't even see them with your eyes. They can only live by hiding inside other things, like our bodies. When they get inside, they use the body's stuff to make more tiny ninjas! These new ninjas can then make even more! Luckily, our bodies can usually fight them off and make us feel better again."

Meta Llama 3.1 70B: "**Germs Called Viruses.** Do you know what's too small to see? **Viruses**! They're like tiny, tiny bugs that can make us sick. A virus has a special coat around it. It needs a host (like our body) to make more of itself. When it's inside us, it can make us feel icky. **But don't worry!** We can stay healthy by washing our hands. Doctors and scientists work hard to keep us safe from viruses. **You're a superstar for learning about viruses!**"

### 4.2 Hypothesis 1

The first hypothesis was that GPT-4 will be the highest-rated LLM among nursery teachers for generating child-appropriate scientific content.

Surprisingly, the GPT-4 model got the lowest ratings of all LLMs in this survey (see fig. 1). Claude was rated the best, followed by Gemini and Llama. A Friedman test revealed significant differences between the ratings of the four LLMs, $\chi^2$ (df = 3, N = 30) = Y, $p <$ 0.001, with post hoc pairwise comparisons (Bonferroni-corrected) showing that all LLMs were rated significantly different (adjustet $p <$ 0.001). Only Llama and Gemini showed no significant differences (adjusted $p$ = 0.08). Therefore, Hypothesis 1 was not supported, as GPT-4 received significantly lower ratings than the other LLMs for generating child-appropriate scientific content.

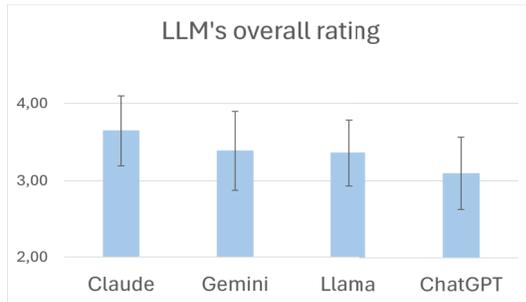

**Figure 1: Aggregated overall ranking of the LLMs.**

### 4.3 Hypothesis 2

The second hypothesis states that there will be significant differences in nursery teachers' ratings for LLM-generated content across scientific disciplines (biology, chemistry, and physics).

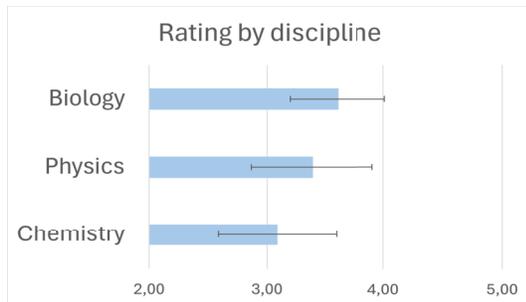

**Figure 2: Aggregated overall rating by discipline.**

As assumed, the results of the LLMs for scientific biological texts are rated best followed by physics and chemistry (see fig. 2). The ratings for each discipline were found to be significantly different from all others (Friedman test: $\chi^2$(df = 2, N = 30) = Y, $p <$ 0.001; all Bonferroni-corrected pairwise comparisons: $p <$ 0.001). Therefore, Hypothesis 2 was supported, with biological content receiving significantly higher ratings than physics and chemistry content.

### 4.4 Hypothesis 3

The third hypothesis suggests that there will be significant differences in nursery teachers' ratings for LLMs across the four evaluation criteria (comprehensibility, language, interest generation, and real-life relatedness).

The nursery teachers were most satisfied with the LLM's real-life relations of the texts followed by appropriate language, comprehensibility, and interest-sparking (see fig. 3). A Friedman test indicated significant variations in the ratings among the four criteria, $\chi^2$(df = 3, N = 30) = Y, $p <$ 0.001. Post hoc pairwise comparisons with Bonferroni correction revealed that not all criteria differ significantly from each other. Language differed significantly from comprehensibility and interest (adjusted $p <$ 0.001). From relation, language differs on a lower significance level (adjusted $p <$ 0.01). Interest and relation differ on an even lower significance level (adjusted $p <$ 0.05), whereas the other two groups (comprehensibility with relation as well as interest) do not show significant differences. Therefore, Hypothesis 3 was supported, with real-life relatedness and language receiving significantly higher ratings than interest generation, while comprehensibility showed mixed results.

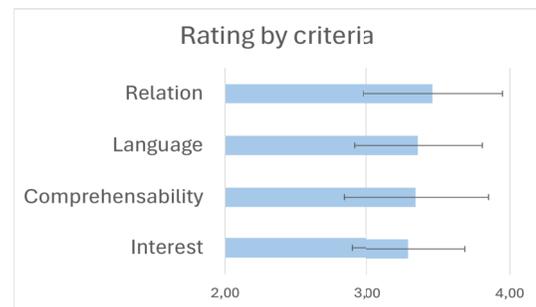

**Figure 3: Aggregated overall rating by criteria.**

## 5 Discussion

This study investigated the capabilities of four leading LLMs in generating child-appropriate scientific content for preschool education, revealing several unexpected findings that challenge existing assumptions about LLM performance in educational contexts. The results provide important insights for both educational technology development and ECSE.

Contrary to our first hypothesis, GPT-4 received the lowest ratings among the evaluated LLMs, with Claude achieving the highest scores followed by Gemini and Llama. This finding contradicts the dominant focus on GPT models in educational LLM research, where over 90% of studies examine OpenAI's GPT capabilities [9]. This unexpected result highlights the importance of systematic comparative evaluations rather than relying on market prominence or general-purpose benchmarks when selecting LLMs for educational applications. Educators and developers should consider using Claude for scientific content generation. Our finding that GPT-4 received the lowest ratings must be understood within our specific research parameters. This result may be influenced by our concise prompting approach, the unique challenges of early childhood science education, and evaluation by nursery teachers with specialized developmental priorities. These factors limit broad generalization to other educational contexts.

The significant differences in performance across scientific disciplines support our second hypothesis and reveal important patterns in LLMs' capabilities. For LLMs, biological topics proved most amenable to child-appropriate translation, likely because they connect directly to children's experiences with their own bodies and observable living things. This stronger performance in biology aligns with previous findings that biological topics often inherit more concrete and observable phenomena compared to other natural sciences [12]. Physical concepts, while somewhat abstract, could often be illustrated through tangible examples that children encounter in play (such as motion and simple machines), resulting in moderate ratings. The lower ratings for chemistry content reflect the inherent challenges in making molecular-level concepts accessible to preschoolers, as they frequently involve microscopic phenomena and abstract concepts that are difficult to translate into concrete experiences meaningful to young children. These findings can guide both content development strategies and curriculum planning in ECSE, suggesting that future AI development should prioritize connecting abstract concepts to observable phenomena within children's immediate experience. The varying performance across disciplines indicates the need for tailored approaches to content generation, with additional support and refinement needed for more abstract topics like chemistry.

Our analysis of evaluation criteria revealed nuanced patterns in LLM-generated content. While statistical differences emerged between criteria ratings, these may reflect challenges in operationalizing distinct aspects of child-appropriate content rather than meaningful variations in LLM capabilities. LLMs demonstrated stronger performance in real-life relatedness and language appropriateness, suggesting reasonable competency in creating foundational child-accessible content. However, the lower ratings for interest generation highlight a significant gap in creating truly engaging scientific content for young children, consistent with Bhandari and Brennan's [5] observations about limitations in AI-generated children's stories. These findings identify specific strengths to leverage and weaknesses to address when implementing LLMs in educational settings. The high ratings for real-life relations and age-appropriate language demonstrate successful implementation of key early childhood education principles [8], while suggesting that current LLM-generated content may need supplementation with additional engaging elements for effective learning experiences.

This study represents an important step in understanding the potential of LLMs in ECSE. While the results demonstrate promising capabilities in generating age-appropriate content, they also highlight significant areas for improvement, particularly in creating engaging material and handling abstract concepts. The findings suggest that LLMs can serve as valuable tools for educators. Our research recognizes that translating scientific concepts for young learners is a core professional competency of educators. Rather than suggesting this task exceeds educators' capabilities, we evaluated how AI systems might support science education by generating content that meets high pedagogical standards. As these models continue to develop, regular comparative evaluations will be crucial for identifying the most effective tools for educational content generation.

While we focused primarily on educator perspectives, parental involvement represents another crucial dimension in early childhood science education. Recent literature indicates parents hold mixed perspectives on AI in educational settings. They recognize potential benefits such as personalized learning experiences [16] but also express concerns about privacy and technology interference with parent-child interactions [4, 10]. Future research should explore how LLM-generated content could support science learning activities at home, creating connections between formal educational settings and home environments.

In our planned follow-up study, we will extend our quantitative evaluation study to include more teachers and also parents. For a mixed methods approach, we also plan to conduct qualitative interviews with both teachers and parents to gain nuances of effective science content for preschool children. This expansion addresses initial informal feedback from nursery teachers indicating that none of the generated texts fully met child-appropriateness standards. By addressing these considerations in future research, we can develop more refined understanding of how LLMs can best serve as tools to support early childhood science education while respecting the essential roles of both educators and parents in children's learning journeys.

## References

[1] Said Al Faraby, Ade Romadhony, and Adiwijaya. 2024. Analysis of LLMs for educational question classification and generation. *Computers and Education: Artificial Intelligence* 7 (2024), 100298. doi:10.1016/j.caeai.2024.100298
[2] Sumit Asthana, Hannah Rashkin, Elizabeth Clark, Fantine Huot, and Mirella Lapata. 28.1. Evaluating LLMs for Targeted Concept Simplification for Domain-Specific Texts. http://arxiv.org/pdf/2410.20763
[3] Jihye Bae, Margaret Shavlik, Christine E. Shatrowsky, Catherine A. Haden, and Amy E. Booth. 2023. Predicting grade school scientific literacy from aspects of the early home science environment. *Frontiers in psychology* 14 (2023), 1113196. doi:10.3389/fpsyg.2023.1113196
[4] Ruqia Safdar Bajwa, Asma Yunus, Hina Saeed, and Asia Zulfqar. 2024. Parenting in the Age of Artificial Intelligence. In *Exploring Youth Studies in the Age of AI*, D.B.A.Mehdi Khosrow-Pour, Zeinab Zaremohzzabieh, Rusli Abdullah, and Seyedali Ahrari (Eds.). IGI Global, 45–68. doi:10.4018/979-8-3693-3350-1.ch003
[5] Prabin Bhandari and Hannah Marie Brennan. 26.0. Trustworthiness of Children Stories Generated by Large Language Models. http://arxiv.org/pdf/2308.00073v1
[6] Theodore L. Brown, Harold Eugene LeMay, Bruce Edward Bursten, and Catherine J. Murphy. 2018. *Chemistry: The central science* (14th edition in si units, global edition ed.). Pearson, Harlow. https://ebookcentral.proquest.com/lib/kxp/detail.action?docID=5832823
[7] Neil A. Campbell, Lisa A. Urry, Michael L. Cain, Steven A. Wasserman, Peter V. Minorsky, and Jane B. Reece. 2021. *Biology: A global approach, global edition* (twelfth edition, global edition ed.). Pearson, Harlow. https://elibrary.pearson.de/book/99.150005/9781292341699
[8] Bilge Nur Doğan and Birkan Güldenoğlu. 2014. Development of Principle of Suitability for Children Scale. *Procedia - Social and Behavioral Sciences* 116 (2014), 3054–3059. doi:10.1016/j.sbspro.2014.01.706
[9] Bingyu Dong, Jie Bai, Tao Xu, and Yun Zhou. 2024. Large Language Models in Education: A Systematic Review. In *2024 6th International Conference on Computer Science and Technologies in Education (CSTE)*. IEEE, 131–134. doi:10.1109/CSTE62025.2024.00031
[10] Jill Glassman, Kathryn Humphreys, Serena Yeung, Michelle Smith, Adam Jauregui, Arnold Milstein, and Lee Sanders. 2021. Parents' Perspectives on Using Artificial Intelligence to Reduce Technology Interference During Early Childhood: Cross-sectional Online Survey. *Journal of medical Internet research* 23, 3 (2021), e19461. doi:10.2196/19461
[11] David Halliday, Robert Resnick, and Jearl Walker. 2013. *Fundamentals of physics* (10th ed. ed.). Wiley. https://permalink.obvsg.at/
[12] Soo-Young Hong and Karen E. Diamond. 2012. Two approaches to teaching young children science concepts, vocabulary, and scientific problem-solving skills. *Early Childhood Research Quarterly* 27, 2 (2012), 295–305. doi:10.1016/j.ecresq.2011.09.006
[13] Chieh-Yang Huang, Jing Wei, and Ting-Hao Kenneth Huang. 2024. Generating Educational Materials with Different Levels of Readability using LLMs. In *Proceedings of the Third Workshop on Intelligent and Interactive Writing Assistants*. ACM, New York, NY, USA, 16–22. doi:10.1145/3690712.3690718
[14] Jana Kähler, Inga Hahn, and Olaf Köller. 2020. The development of early scientific literacy gaps in kindergarten children. *International Journal of Science Education* 42, 12 (2020), 1988–2007. doi:10.1080/09500693.2020.1808908
[15] Pranjal Kumar. 2024. Large language models (LLMs): survey, technical frameworks, and future challenges. *Artificial Intelligence Review* 57, 10 (2024), 1–51. doi:10.1007/s10462-024-10888-y

[16] Pauldy Cornelia Johanna Otermans, Stephanie Baines, Chelsea Livingstone, Monica Pereira, and Dev Aditya. 2024. Chatting with the Future: A Comprehensive Exploration of Parents' Perspectives on Conversational AI Implementation in Children's Education. *International Journal of Technology in Education* 7, 3 (2024), 573–586. doi:10.46328/ijte.812
[17] Konstantinos Ravanis. 2022. Research Trends and Development Perspectives in Early Childhood Science Education: An Overview. *Education Sciences* 12, 7 (2022), 456. doi:10.3390/educsci12070456
[18] Konstantinos Ravanis and George Bagakis. 1998. Science Education in Kindergarten: Sociocognitive perspective. *International Journal of Early Years Education* 6, 3 (1998), 315–327. doi:10.1080/0966976980060306
[19] Nancy Butler Songer and Amelia Wenk Gotwals. 2012. Guiding explanation construction by children at the entry points of learning progressions. *Journal of Research in Science Teaching* 49, 2 (2012), 141–165. doi:10.1002/tea.20454
[20] Jennifer Weber, Maria Valentini, Téa Wright, Katharina von der Wense, and Eliana Colunga. 2024. Evaluating LLMs as Tools to Support Early Vocabulary Learning. *Proceedings of the Annual Meeting of the Cognitive Science Society* 46 (2024), 2633.
[21] Ziyu Yang. 2024. *Enhancing the Comprehension: Text Simplification Approaches and the Role of Large Language Models*. Ph. D. Dissertation. Temple University, Philadelphia, USA. doi:10.34944/DSPACE/10208
[22] Barbara Zelger and Simone Stefan. 2020. Die Festlegung von qualitativen Entscheidungskriterien mit GABEK® am Beispiel von Kinderliteratur. In *Symposium Qualitative Sozialforschung 2019*, Margit Raich, Julia Müller-Seeger, and Helmut Ebert (Eds.). Hallesche Schriften zur Betriebswirtschaft, Vol. 35. Springer Fachmedien Wiesbaden, Wiesbaden, 243–258. doi:10.1007/978-3-658-32463-6{_}13